\documentstyle[prb,aps,epsf]{revtex}
\begin{document}
\draft
\author{C. Bagnuls\thanks{%
e-mail: bagnuls@spec.saclay.cea.fr}} 
\address{Service de Physique de l'Etat Condens\'{e}, C. E. Saclay, F91191 Gif-sur-Yvette Cedex, France}

\author{ 
C. Bervillier\thanks{%
e-mail: bervil@spht.saclay.cea.fr} }
\address{Service de Physique Th\'{e}orique, C. E. Saclay, F91191 Gif-sur-Yvette Cedex, France}

\author{D. I. Meiron\thanks{e-mail: dim@its.caltech.edu 
}}
\address{Applied and Computational Mathematics,
MC 08-31, Caltech, Pasadena, CA 91125, USA
}
\author{B. G. Nickel\thanks{e-mail: bgn@physics.uoguelph.ca
}}
\address{Department of Physics, University of Guelph, Guelph, Ontario, Canada N1G2W1}
\title{Addendum-erratum to:\\ ``{\em Nonasymptotic critical behavior from field
theory at }$d=3${\em . II. The ordered-phase case}''\\
Phys. Rev. B35, 3585 (1987).}
\date{January 11, 2002}
\maketitle

\begin{abstract}
This note is intended to emphasize the existence of estimated Feynman
integrals in three dimensions for the free energy of the $O(1)$ scalar
theory up to five loops which may be useful for other work. We also correct
some errors in the original paper.
\end{abstract}
\paragraph*{Addendum}
In tables  \ref{corrtab1} and \ref{corrtab2}  we present the corrected table I of the original paper \cite{736}. They
display the values of the Feynman integrals of the $O(1)$ scalar theory
contributing to the free energy up to five loops and their symmetry factors.
The values differ from those Feynman integrals previously calculated in  Ref.\ \onlinecite
{323} due to the necessity of
introducing a ``soft'' mass parameter instead of the usual renormalized (at
zero-momentum) mass (see the text of the original paper \cite{736} and  Ref.\ \onlinecite
{733} for details). Consequently, many of the estimates of Feynman integrals
presented in table I of   Ref.\ \onlinecite{736} have been extracted from  Ref.\ \onlinecite{323} by accounting for a
harmless 3-d renormalization in order to get a soft-mass parameter
(characterizing a minimal subtraction scheme similar to that introduced in 
 Ref.\ \onlinecite{2272}). Since the five-loop contributions to the free energy involve $%
\varphi ^{3}$ vertices mixed to $\varphi ^{4}$ vertices, Feynman integrals
which are different or cannot be obtained from those considered
in  Ref.\ \onlinecite{323} have been estimated in three dimensions for the occasion \cite{foot1}.
These are:

\begin{itemize}
\item  the four-loop integrals with Heap's numbers 13--17 (column $h$ in the
following tables \ref{corrtab1} and \ref{corrtab2}, see table IV of Ref.\ \onlinecite{4718}). They involve only $\varphi
^{3}$ vertices. They have been estimated and successfully compared to
similar calculations extracted from Ref.\ \onlinecite{362}.

\item  the five-loop integrals with Heap's numbers 80--102 (see table V of 
Ref.\ \onlinecite{4718}). They involve $\varphi ^{3}$ vertices mixed with a single $%
\varphi ^{4}$ vertex. They have been calculated exclusively for this work.

\item  the five-loop integrals with Heap's numbers 103--118 (see table V of 
Ref.\ \onlinecite{4718}). They involve only $\varphi ^{3}$ vertices. They have been
calculated for this work and successfully compared to similar calculations
extracted from Ref.\ \onlinecite{4721}.
\end{itemize}

\paragraph*{Errata}

Tables \ref{corrtab1} and \ref{corrtab2} (corrected table I of  Ref.\ \onlinecite{736})
account for two corrections compared to the original
paper \cite{736}:

\begin{itemize}
\item  for $b=5$, $h=15$, $l=0$, $m=1$, in the column ``Value at $d=3$'',
one reads $6.24797746$ instead of $4.21825152$.

\item  Eq. (A13), which gives the contributions $b=5$, $h=49$, $l=2$, $m=1$
and $b=5$, $h=50$, $l=2$, $m=2$ in table I, should read: 
\begin{equation}
4I_{1}^{(5,2)}(1)+2I_{2}^{(5,2)}(1)=-0.1202442510-1.558293723\ln \left(
r_{0}^{\prime }/g_{0}^{2}\right)   \eqnum{A13}
\end{equation}
\end{itemize}

As consequences, in table II of   Ref.\ \onlinecite{736}, the values of $F_{500}$ and $F_{510}$ should read:

\begin{eqnarray*}
F_{500} &=&-0.45163891229\times 10^{-7} \\
F_{510} &=&0.80687809188\times 10^{-7}
\end{eqnarray*}
in agreement with table 1 of  Ref.\ \onlinecite{2376}. In addition, as indicated in the note [12] of  Ref.\ \onlinecite{4073}, the five loop coefficients of
$X(v)$ and $F(v)$ given in table III of Ref.\ \onlinecite{736} should be modified but also that of
$\tilde F(v)$, so that it is preferable to redone table III (see the present table \ref{corrtab3}).

Moreover, the following equations were not correctly written, they should
read:

\begin{equation}
\Gamma _{0,-}^{(0,0)}\left( r_{0},g_{0},M_{0},\epsilon \right) =\frac{1}{2}%
r_{0}M_{0}^{2}+g_{0}\frac{M_{0}^{4}}{24}+\sum_{b=1}^{5}\Gamma _{b}\left(
r_{0},g_{0},M_{0},\epsilon \right)  \eqnum{3.1}
\end{equation}

\begin{equation}
H_{blk}=\left( 2-\frac{b}{2}-l\right) F_{blk}+\left( l+1\right) F_{b,l+1,k}-%
\frac{1}{2}\left( k+1\right) F_{b,l,k+1}  \eqnum{3.20}
\end{equation}

\begin{equation}
Q_{blk}=2\left( 2-\frac{b}{2}-l\right) H_{b,l-1,k}+\left( 1+2l\right)
H_{blk}-\left( k+1\right) H_{b,l-1,k+1}  \eqnum{3.22}
\end{equation}

To get  the value of $A^{+}/A^{-}$ given in table VI of  Ref.\ \onlinecite{736} using Eq.\ (4.9), the values of $\tilde{F}(v^{*})$ given in  table V 
must be divided by  $v^{*}$.

In table VII of   Ref.\ \onlinecite{736}, the values of $Y_{3}$ should read: 
\begin{eqnarray*}
&&-2.07825\times 10^{-3} \\
&&3.45588\times 10^{-2}
\end{eqnarray*}

\begin{figure}
   \centerline{ \epsfxsize=12in \epsfbox[-150 100 750 700]{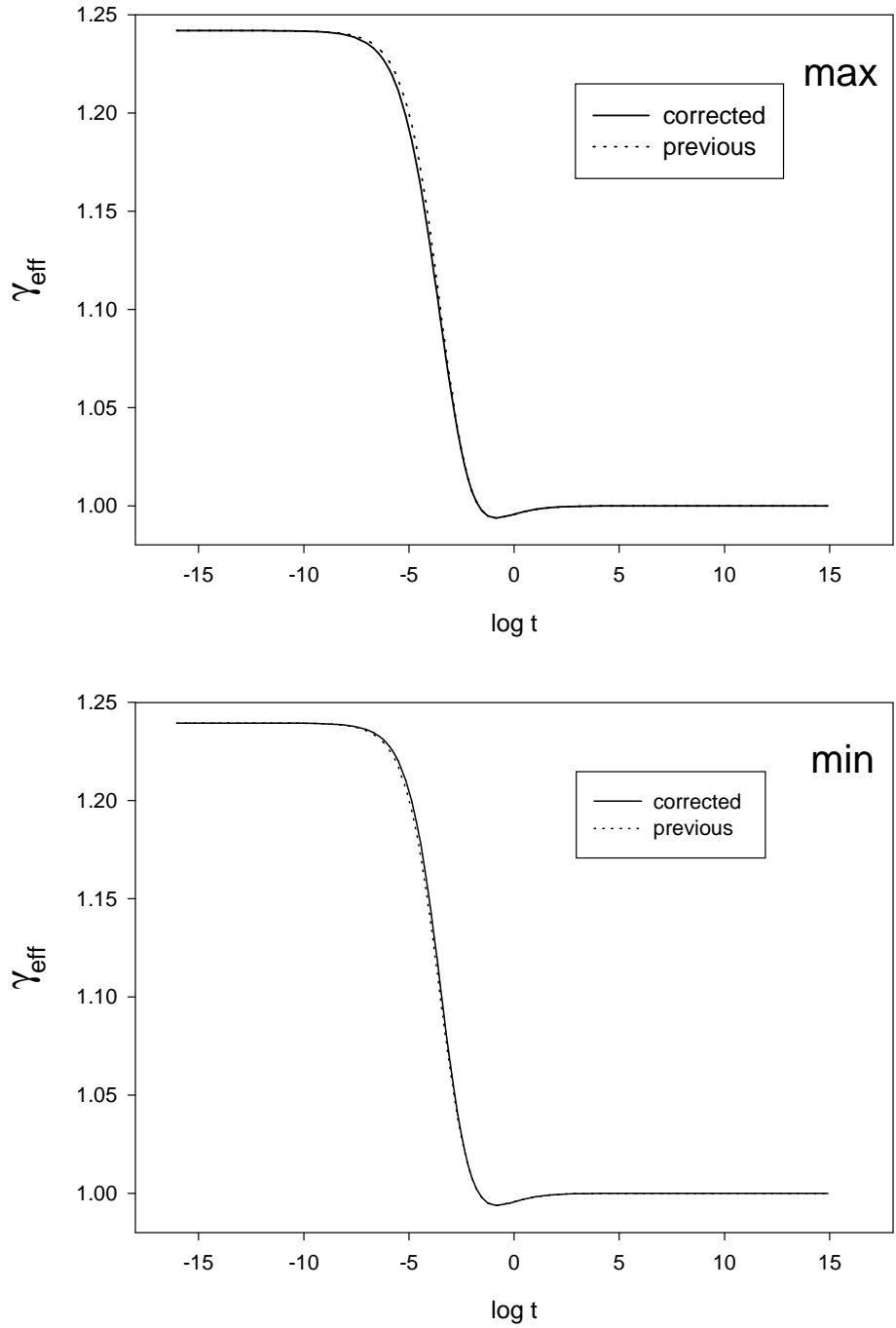} }
\caption{
Illustrations on the effective exponent $\gamma_{eff}(t)$ of the (small) effect of the corrections
of the errors mentionned in the text. The two uncertainty bounds ``max'' and ``min'' corresponding
to the resummation criteria of  Ref.\ \protect\onlinecite{736} are
displayed. These curves have been obtained using the results of  Ref.\ \protect\onlinecite{5187}.
}
\label{fig1}
\end{figure}

All these corrections have been accounted for in  Ref.\ \onlinecite{5187} where 
an updated calculation of the complete critical-to-classical crossover is presented.
In addition, 
we have explicitly
verified (see fig. \ref{fig1}) that the errors have had no important consequence on the final
results as it could be clearly deduced from a comparison of our
estimates of universal amplitude-combinations \cite{736} with those of Guida
and Zinn-Justin \cite{3639} who used the corrected series.

\paragraph*{Acknowledgements}

We thank Prof. H. Kleinert for having incited us to publish, at last, the
corrected table I of Feynman graphs.

\newpage

% \begin{figure}
% \caption{}
% \label{}
% \end{figure}

\newpage

\begin{table}
\caption{Corrected table I. 
See the caption in the original paper Ref.\ \protect\onlinecite{736}. \label{corrtab1}
}
\begin{tabular}{crrrrcd|cllllld}

$b$ & $h$ & $l$ & $m$ & P$_{m}^{-1}$ &  & Value at $d=3$ & $b$ & $h$ & $l$ & 
$m$ & P$_{m}^{-1}$ &  & Value at $d=3$ \\ \hline
1 & --- & 0 & 1 & 2 &  & $-$4/3 &  &  &  &  &  &  &  \\ 
2 & 1 & 1 & 1 & 12 &  & $-2\ln \left( y\right) $ & 5 & 49 & 2 & 1 & 48 &  & 
\\ 
&  &  &  &  &  &  &  & 50 &  & 2 & 24 &  & See Eq. (A.13) \\ 
3 & 1 & 0 & 1 & 48 &  & $-$22.79417368 &  & 51 &  & 3 & 24 &  & $%
-$0.007497124  \\ 
&  &  &  &  &  & $+16\ln \left( y\right) $ &  &  &  &  &  &  & $%
-$0.086950305$\ln \left( y\right) $ \\ 
& 2 & 1 & 1 & 8 &  & 4.107471254 &  & 52 &  & 4 & 32 &  & 0.36927278680
\\ 
& 3 & 2 & 1 & 16 &  & 0.5194312413 &  & 53 &  & 5 & 64 &  & 0.36927278680
\\ 
& 4 &  & 2 & 24 &  & 0.17390061070 &  & 54 &  & 6 & 8 &  & 0.22602937610
\\ 
&  &  &  &  &  &  &  & 55 &  & 7 & 8 &  & 0.26394187370 \\ 
4 & 3 & 0 & 1 & 48 &  & $-$19.73920880$\ln \left( y\right) $ &  & 56 &  & 8 & 8
&  & 0.97910169300$\times 10^{-1}$ \\ 
& 5 & 1 & 1 & 24 &  & $-$0.2964527240 &  & 57 &  & 9 & 32 &  & 
0.29097562780  \\ 
&  &  &  &  &  & $-\left( 4/3\right) \ln \left( y\right) $ &  & 58 &  & 10 & 
8 &  & 0.26394187370 \\ 
& 6 &  & 2 & 16 &  & 2.0657193571 &  & 59 &  & 11 & 16 &  & 0.22602937610
\\ 
& 7 &  & 3 & 8 &  & 1.7234905497 &  & 60 &  & 12 & 16 &  & 0.22294544960
\\ 
& 8 &  & 4 & 8 &  & 1.2405960978 &  & 61 &  & 13 & 4 &  & 0.16404726530
\\ 
&  &  &  &  &  &  &  & 62 &  & 14 & 16 &  & 0.19621789690 \\ 
& 9 & 2 & 1 & 16 &  & 0.43051311360 &  & 63 &  & 15 & 8 &  & 
0.94895087300$\times 10^{-1}$ \\ 
& 10 &  & 2 & 4 &  & 0.31160313040 &  & 64 &  & 16 & 8 &  & 0.18956812860
\\ 
& 11 &  & 3 & 4 &  & 0.12578653970 &  & 65 &  & 17 & 8 &  & 
0.93486460600$\times 10^{-1}$ \\ 
& 12 &  & 4 & 8 &  & 0.79516908900$\times 10^{-1}$ &  & 66 &  & 18 & 4 &  & 
0.94895087300$\times 10^{-1}$ \\ 
&  &  &  &  &  &  &  & 67 &  & 19 & 16 &  & 0.97910169300$\times 10^{-1}$ \\ 
& 13 & 3 & 1 & 48 &  & 0.18361624610 &  & 68 &  & 20 & 8 &  & 
0.18956812860 \\ 
& 14 &  & 2 & 16 &  & 0.80721242900$\times 10^{-1}$ &  & 69 &  & 21 & 4 &  & 
0.66591760000$\times 10^{-1}$ \\ 
& 15 &  & 3 & 8 &  & 0.37859728700$\times 10^{-1}$ &  & 70 &  & 22 & 8 &  & %
0.53216192900$\times 10^{-1}$ \\ 
& 16 &  & 4 & 12 &  & 0.14620245800$\times 10^{-1}$ &  & 71 &  & 23 & 16 & 
& 0.49947540000$\times 10^{-1}$ \\ 
& 17 &  & 5 & 72 &  & 0.12244670000$\times 10^{-1}$ &  & 72 &  & 24 & 4 &  & 
0.17156245110 \\ 
&  &  &  &  &  &  &  & 73 &  & 25 & 4 &  & 0.72746695800$\times 10^{-1}$ \\ 
5 & 15 & 0 & 1 & 144 &  & 6.24797746 &  & 74 &  & 26 & 2 &  & %
0.59161184100$\times 10^{-1}$ \\ 
&  &  &  &  &  & $+$4.602913152$\ln \left( y\right) $ &  & 75 &  & 27 & 4 &  & 
0.56505190000$\times 10^{-1}$ \\ 
&  &  &  &  &  & $+$4$\ln ^{2}\left( y\right) $ &  & 76 &  & 28 & 8 &  & %
0.89556123600$\times 10^{-1}$ \\ 
& 16 &  & 2 & 128 &  & 22.90931839 &  & 77 &  & 29 & 12 &  & %
0.32950940000$\times 10^{-1}$ \\ 
& 17 &  & 3 & 32 &  & 16.60229522 &  & 78 &  & 30 & 2 &  & %
0.34817630000$\times 10^{-1}$ \\ 
&  &  &  &  &  &  &  & 79 &  & 31 & 16 &  & 0.34355390400$\times 10^{-1}$ \\ 
& 31 & 1 & 1 & 48 &  & $-$0.142393552 &  &  &  &  &  &  &  \\ 
&  &  &  &  &  & $-$1.038862482$\ln \left( y\right) $ &  & 80 & 3 & 1 & 32 & 
& 0.16930103539 \\ 
& 32 &  & 2 & 12 &  & $-$0.122911141 &  & 81 &  & 2 & 16 &  & %
0.70750283580$\times 10^{-1}$ \\ 
&  &  &  &  &  & $-$0.767152192$\ln \left( y\right) $ &  & 82 &  & 3 & 16 &  & 
0.33556467810$\times 10^{-1}$ \\ 
& 33 &  & 3 & 32 &  & 1.373092004 &  & 83 &  & 4 & 16 &  & %
0.67245000060$\times 10^{-1}$ \\ 
& 34 &  & 4 & 16 &  & 2.449689513 &  & 84 &  & 5 & 8 &  & %
0.70750283580$\times 10^{-1}$ \\ 
& 35 &  & 5 & 8 &  & 1.072299357 &  & 85 &  & 6 & 8 &  & 0.11737660585
\\ 
& 36 &  & 6 & 16 &  & 0.629580783 &  & 86 &  & 7 & 4 &  & %
0.30521594800$\times 10^{-1}$ \\ 
& 37 &  & 7 & 8 &  & 0.785125191 &  & 87 &  & 8 & 8 &  & %
0.33556467810$\times 10^{-1}$ \\ 
& 38 &  & 8 & 4 &  & 0.729050922 &  & 88 &  & 9 & 4 &  & %
0.50133149830$\times 10^{-1}$ \\ 
& 39 &  & 9 & 8 &  & 0.853563223 &  & 89 &  & 10 & 8 &  & %
0.29350397320$\times 10^{-1}$ \\ 
& 40 &  & 10 & 4 &  & 0.489725124 &  & 90 &  & 11 & 4 &  & %
0.30521594800$\times 10^{-1}$ \\ 
& 41 &  & 11 & 12 &  & 0.31815728 &  & 91 &  & 12 & 4 &  & %
0.23674907200$\times 10^{-1}$ 
\end{tabular}
\end{table}
\newpage
\begin{table}
\caption{Corrected table I continued. \label{corrtab2}}
\begin{tabular}{crrrrcd|cllllld}
$b$ & $h$ & $l$ & $m$ & P$_{m}^{-1}$ &  & Value at $d=3$ & $b$ & $h$ & $l$ & 
$m$ & P$_{m}^{-1}$ &  & Value at $d=3$ \\ \hline
5 & 92 & 2 & 13 & 4 &  & \multicolumn{1}{l|}{0.12127180970$\times 10^{-1}$}
& 5 & 105 & 4 & 3 & 96 &  & 0.27143831355$\times 10^{-1}$ \\ 
& 93 &  & 14 & 2 &  & \multicolumn{1}{l|}{0.11619507150$\times 10^{-1}$} & 
& 106 &  & 4 & 16 &  & 0.16765892110$\times 10^{-1}$ \\ 
& 94 &  & 15 & 8 &  & \multicolumn{1}{l|}{0.99110117700$\times 10^{-2}$} & 
& 107 &  & 5 & 8 &  & 0.11447149611$\times 10^{-1}$ \\ 
& 95 &  & 16 & 8 &  & \multicolumn{1}{l|}{0.38967085760$\times 10^{-1}$} & 
& 108 &  & 6 & 16 &  & 0.13776551025$\times 10^{-1}$ \\ 
& 96 &  & 17 & 4 &  & \multicolumn{1}{l|}{0.17420445300$\times 10^{-1}$} & 
& 109 &  & 7 & 32 &  & 0.10033568785$\times 10^{-1}$ \\ 
& 97 &  & 18 & 4 &  & \multicolumn{1}{l|}{0.16062044550$\times 10^{-1}$} & 
& 110 &  & 8 & 8 &  & 0.65847074272$\times 10^{-2}$ \\ 
& 98 &  & 19 & 4 &  & \multicolumn{1}{l|}{0.18506772840$\times 10^{-1}$} & 
& 111 &  & 9 & 4 &  & 0.41535270157$\times 10^{-2}$ \\ 
& 99 &  & 20 & 2 &  & \multicolumn{1}{l|}{0.68328380000$\times 10^{-2}$} & 
& 112 &  & 10 & 8 &  & 0.47864827139$\times 10^{-2}$ \\ 
& 100 &  & 21 & 8 &  & \multicolumn{1}{l|}{0.79486181900$\times 10^{-2}$} & 
& 113 &  & 11 & 16 &  & 0.36515474403$\times 10^{-2}$ \\ 
& 101 &  & 22 & 4 &  & \multicolumn{1}{l|}{0.57948451700$\times 10^{-2}$} & 
& 114 &  & 12 & 16 &  & 0.31588757411$\times 10^{-2}$ \\ 
& 102 &  & 23 & 4 &  & \multicolumn{1}{l|}{0.54193873300$\times 10^{-2}$} & 
& 115 &  & 13 & 4 &  & 0.16563666713$\times 10^{-2}$ \\ 
&  &  &  &  &  & \multicolumn{1}{l|}{} &  & 116 &  & 14 & 12 &  & %
0.13970600340$\times 10^{-2}$ \\ 
& 103 & 4 & 1 & 128 &  & \multicolumn{1}{l|}{0.99256755397$\times 10^{-1}$}
&  & 117 &  & 15 & 48 &  & 0.12151455340$\times 10^{-2}$ \\ 
& 104 &  & 2 & 16 &  & \multicolumn{1}{l|}{0.34440788678$\times 10^{-1}$} & 
\multicolumn{1}{l}{} & 118 &  & 16 & 16 &  & 0.11715681490$\times 10^{-2}$
\end{tabular}
\end{table}

\begin{table}
\caption{Modified table III of  Ref.\ \protect\onlinecite{736}. \label{corrtab3}}

\begin{tabular}{dddd}
$X$ & $S$ & $\tilde{F}$ & $F$ \\  \hline
1.5 & 1.0 & 4.5 & $-$0.25 \\ 
0.0 & 0.0 & 0.0 & 0.0 \\ 
0.0 & $-$0.16666666 & 0.33333333 & $-$7.1299755$\times $10$^{-3}$
\\ 
2.9928535$\times $10$^{-2}$ & 2.9018571$\times $10$^{-2}$ & $-$%
0.17775128 & 3.8703438$\times $10$^{-3}$ \\ 
$-$2.3069166$\times $10$^{-2}$ & 1.6869642$\times $10$^{-3}$ & 
6.8988036$\times $10$^{-2}$ & $-$3.2942831$\times $10$^{-3}$ \\ 
2.0211486$\times $10$^{-2}$ & 4.3368472$\times $10$^{-2}$ & 
$-$0.12771112 & 4.1253272$\times $10$^{-3}$%
\end{tabular}
\end{table}


\begin{thebibliography}{9}
\bibitem{foot1}
In Ref.\ \onlinecite{323} Feynman graphs with two and four external legs of a pure $%
\varphi ^{4}$ theory (hence, with regard to similar contributions to the
free energy, involving two and four $\varphi ^{3}$ vertices respectively)
have been exclusively considered. Consequently, when the number of $\varphi ^{3}$
vertices exceeds four, new integrals have to be estimated.
\end{thebibliography}
\end{document}